\documentclass[sigchi]{acmart}

\usepackage{tabularx}
\usepackage{blindtext}
\usepackage[utf8]{inputenc}
\usepackage{caption}
\usepackage{subcaption}
\usepackage{hhline}
\usepackage{graphicx}

\AtBeginDocument{%
  \providecommand\BibTeX{{%
    \normalfont B\kern-0.5em{\scshape i\kern-0.25em b}\kern-0.8em\TeX}}}

\setcopyright{rightsretained} 

\begin{document}

\title{Augmentix - An Augmented Reality System for asymmetric Teleteaching}

\author{Nico Feld}
\email{feldn@uni-trier.de}
\affiliation{%
  \institution{University of Trier}
  \streetaddress{Universitätsring 15}
  \city{Trier}
  \state{Germany}
  \postcode{56290}
}

\renewcommand{\shortauthors}{Feld}

\begin{abstract}
Using augmented reality in education is already a common concept, as it has the potential to turn learning into a motivational learning experience. However, current research only covers the students site of learning. Almost no research focuses on the teachers' site and whether augmented reality could potentially improve his/her workflow of teaching the students or not. Many researchers do not differentiate between multiple user roles, like a student and a teacher. To allow investigation into these lacks of research, a teaching system ''Augmentix'' is presented, which includes a differentiation between the two user roles ''teacher'' and ''student'' to potentially enhances the teachers workflow by using augmented reality. In this system's setting the student can explore a virtual city in virtual reality and the teacher can guide him with augmented reality.
\end{abstract}



\keywords{augmented reality, virtual reality, teleteaching, teleeducation, tangibles, collaborative virtual environment}

\maketitle

\section{Introduction}
\label{sec:introduction}
Technologies such as augmented reality (AR) have the potential to enrich the physical world with virtual content by augmenting the physical environment by digital information. This combination of (mixing) a physical with a virtual environment may simplify the interaction with this virtual content as the user still has reference to the physical, known environment \cite{zhou2008trends}.
Since the technology of augmented reality started to rise, it was examined for benefits they can bring to teaching and education. Especially since the rapid improvements in mobile technologies (like smartphones) made them accessible for a wider audience. 

Therefore, one scenario which we think could greatly benefit from using augmented reality would be virtual classrooms. The students can freely move around a virtual environment and the teacher does use AR to guide them through the virtual environment and teach them about a specific topic. This scenario does not only allow distributed users to participate in such a learning experience, but technologies like AR could further improve their experience \cite{sommerauer2018augmented}. 

Furthermore we think that AR could also improve the workflow of the guide. By the combination of the real physical environment and the virtual environment AR provides natural and intuitive interaction, especially with the usage real objects being used to interact with he virtual environment, called ''Tangibles'' \cite{HCI-049}. This could allow the guide to teach the student in a natural and intuitive way, as he/she would do in a real city tour, without having to adapt his/her workflow as much as he would need to with non-AR system.

Considering this scenario, we may define this situation as an asymmetric CVE as defined by Feld \& Weyers \cite{mci/Feld2019} In an asymmetric CVE, they specify two user roles. A primary is a user who controls the CVE and contributes most of the content. 
A secondary on the other hand is a user who mostly consumes this content and only have minor contribution. In an typical asymmetric teleteaching CVE, a teacher would be a primary while the students would be multiple secondaries, as the teacher is providing the information the students are consuming. In a symmetric CVE, every user has the same role and therefore every user contributes equally. A typical symmetric CVE would be a virtual room for autonomous group work.

While Feld \& Weyers only found two examples for symmetric CVEs systems for teaching \cite{Ibanez:jucs_18_15:architecture_for_collaborative_learning,santos2014survey} using AR they found no asymmetric CVE for teaching. The authors of this paper additionally found one system where one user had another role as the others \cite{bower2017collaborative}, but this only included the ability to show slides. Therefore, there is a lack of research regarding asymmetric CVEs with AR the scenario we suggested, where there is one teacher and multiple students.

In addition Feld \& Weyers also found no examples for asymmetric CVE where the primary uses AR to control and contribute to the CVE. Every usage of AR in the research found was limited to the secondaries to turn learning into a motivational learning experience \cite{sommerauer2018augmented}, but never tackles the question if AR could potentially improve their workflow in teaching secondaries.

Therefore we identified two major research questions about asymmetric CVEs using AR for a system implementing the given scenario: 
\begin{enumerate}
    \item How to design an asymmetric CVE in context of education?
    \item Can a primary benefit from AR?
\end{enumerate}

To answer these questions, we present the asymmetric CVE for teaching Augmentix by implementing the suggested scenario. In this system, the primary uses AR to control the CVE and the secondaries are guided by the primary through an virtual environment. In the prototype of Augmentix the secondaries can have a virtual city tour in virtual reality (VR) through an antic version (400 A.D.) of the german city Trier. This city tour then is guided by a primary, who is in his office using AR.

\begin{figure}
    \centering
    \includegraphics[width=0.47\textwidth]{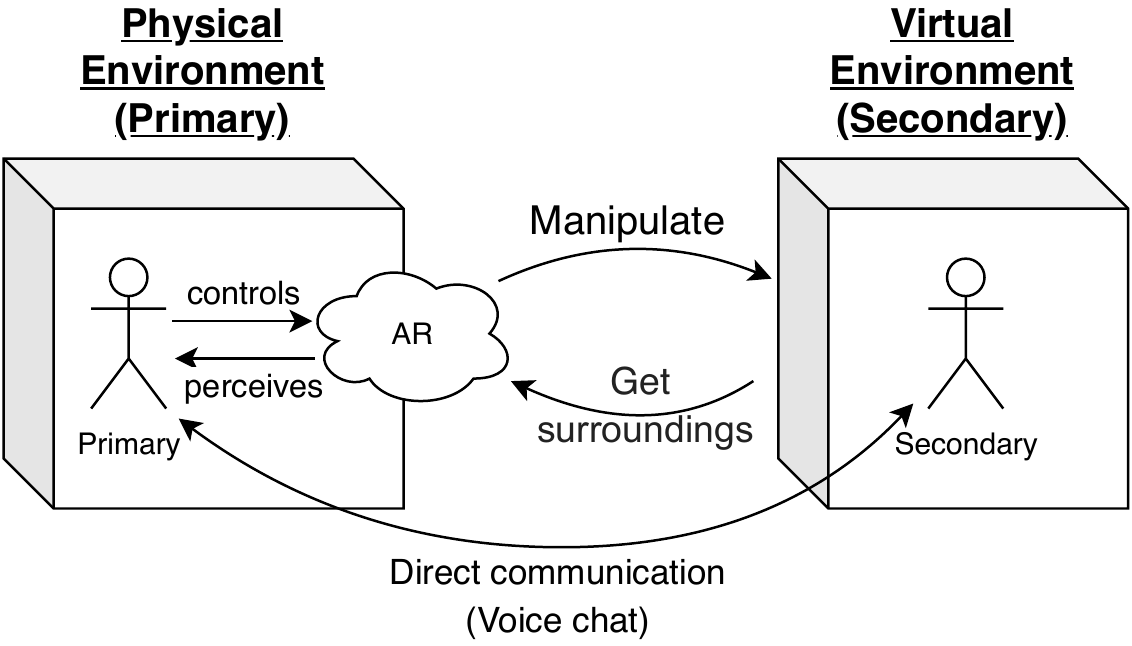}
    \caption{The CVE of Augmentix}
    \label{fig:setup}
\end{figure}

In this paper, CVE describes the whole environment both the primary and the secondaries are in, therefore the primaries virtual and physical environment and the secondaries virtual environment. This CVE is visualized in figure \ref{fig:setup}. The virtual environment of the secondary describes the environment the secondary is in. In the case of Augmentix, this is a virtual representation of the antic Trier. The primaries environment on the other hand is split into his/her physical environment and a virtual environment for augmentation.

To design and implement a system for this scenario, first requirements of this system get derived (section \ref{sec:requirements}). Based on these requirements, a workflow is designed (section \ref{sec:workflow}), which then is implemented in a prototype (section \ref{sec:implementation}). This prototype then was evaluated in a small expert interview (section \ref{sec:interview}). Based on the findings of this interview the results were discussed and concluded (section \ref{sec:conclusion}) and potential future work is defined (section \ref{sec:future}).

\section{Requirements}
\label{sec:requirements}

To implement a prototype for the given scenario, the system must meet certain design requirements (DR). 
Based on these DRs we concluded certain technical requirements (TR). In the following first the DRs are defined and then the resulting TRs are listed.

\subsection{Design requirements}

For this scenario, secondaries need to freely move around the virtual city and the primary can then focus on one secondary to communicate with him and guide him through the city. To do this, the primary must be able to display additionally information to the secondary and manipulating the CVE. To allow the primary to display the correct information, he/she must know the virtual environment of the secondary. Furthermore, the primary needs to also be able to interact with his/her real environment to be able to test possible enhancements in his/her workflow to answer the two identified research questions. Therefore we define the following requirements:
\begin{enumerate}
    \item[DR 1.] Communication
    \begin{enumerate}
        \item[1.] The primary and secondaries need to communicate
        \item[2.] The primary needs to see the secondary and his/her virtual environment
        \item[3.] The primary needs to steer the attention of the secondary towards certain points of interest
    \end{enumerate}
    \item[DR 2.] Interaction
    \begin{enumerate}
        \item[1.] The primary needs to interact and manipulate the CVE
        \item[2.] The primary needs to show additional information to the secondary
        \item[3.] The primary needs to use AR
        \item[4.] The primary needs to potentially switch between multiple secondaries
    \end{enumerate}
    \item[DR 3.] Navigation
    \begin{enumerate}
        \item[1.] The secondary needs to  freely move inside the CVE
        \item[2.] The primary needs to interact with the secondary regardless their position in the environment
    \end{enumerate}
\end{enumerate}

\subsection{Technical requirements}
Based on these DRs we identified the following technical requirements:

\begin{enumerate}
    \item[TR 1] Network Connection between both parties
    \item[TR 2] Communication devices for both parties
    \item[TR 3] AR-Display Device for primary
    \item[TR 4] Input Device for primary
    \item[TR 5] Tangibles for primary
    \item[TR 6] Display \& Input Device for secondary
\end{enumerate}

Because the users are using the same CVE, some sort of network connection between the users is mandatory for this system (TR 1), to synchronize the two local environments. To let the primary and secondaries communicate (DR 1.1) some sort of communication device is needed (TR 2).
To display the virtual environment to the primary (DR 1.2, DR 1.3, DR 2.1, DR 2.2 and DR 3.2) a display device is needed. To further meet DR 2.3 this display device must must support AR (TR 3). These also lead to TR 4 as the primary must also have a way to interact with the CVE. In Augmentix, we decided to use a common smartphone as an AR display and input device. This also meets TR 2 as the built in microphone can be used. In future work this system can be extended to support an AR-HMD (eg. Hololens 2), as this allows an immersive experience and supports full hand tracking, which could further improve the primaries workflow.

To let the primary interact with the CVE (DR 2.1) while using AR (DR 2.3) to interact with his real environment the authors used tangibles (TR 5), as they can be used as an physically-based interface metaphor \cite{billinghurst2008tangible}.

To meet DR 3.1 the secondary needs a display and input to move freely inside the CVE. In this system, he/she uses virtual reality to immerse him-/herself into and interact with the CVE (TR 6). As research showed VR, like AR, can create a motivational learning experience \cite{borst2018fieldtrips} and therefore the authors decided to implement the secondaries site with VR.

\section{Workflow and Implementation}
\label{sec:workflow}
''Augmentix'' was designed with support for multiple secondaries in mind. While the technical implementation only supports one secondary in this prototype, the workflow was defined for multiple secondaries (DR 2.4). 

When the secondary starts Augmentix he/she is directly put into the virtual environment and can start freely exploring via the teleport metaphor. Because the teleport metaphor does only allow to teleport for a short distance, the secondary does have a map of the virtual environment attached to his/her left hand. This map can be shown and hidden via pressing a button and show buttons for every major POI in the virtual environment. By clicking one of those button on the map, the secondary is teleported to this POI. With the short teleport and the teleport to POIs DR 3.1 is fulfilled.

The primary site of the CVE on the other hand was divided into two major parts. The ''Map-Hub'' and the ''Pickup-Shovel''. The Map-Hub should give an overview to the primary about the position of the secondaries and allow him to pick one which he/she wants to further interact with. The primary can return to the Map-Hub at any time for an overview or to focus on another secondary.
To meet DR 2.3 the Map-Hub was realized with AR in mind. Therefore a real map of the antic Trier was augmented and the secondaries are displayed with a arbitrary avatar on their position inside the city. The primary can walk up upon this map and can see all the secondaries and their position in the city.
\begin{figure}
    \centering
    \begin{subfigure}[b]{0.47\textwidth}
        \centering
         \includegraphics[width=\textwidth]{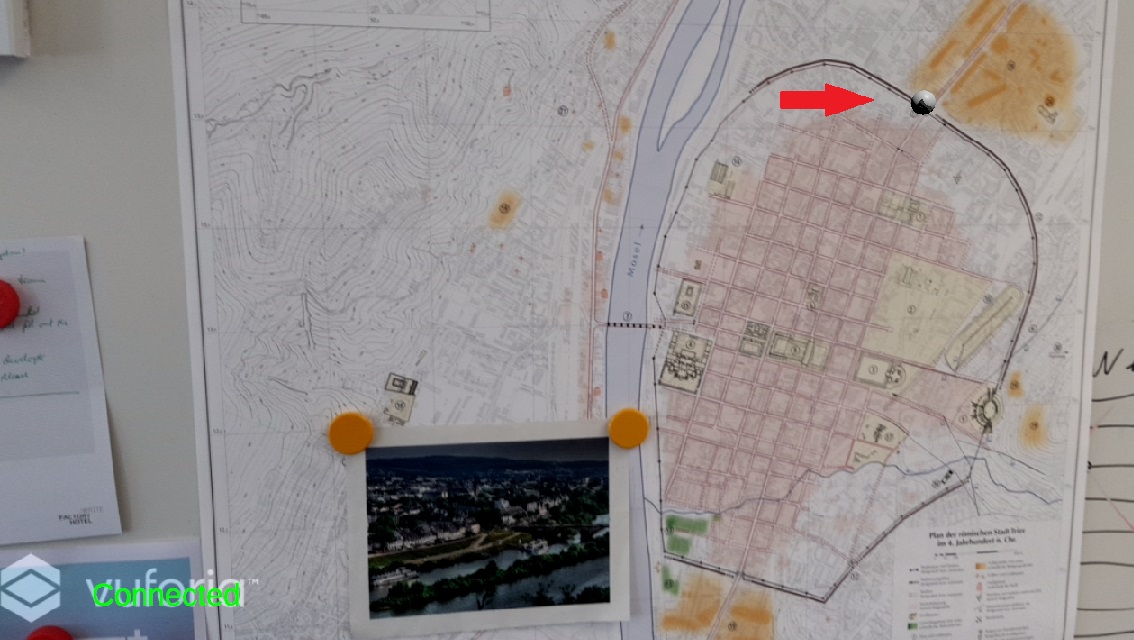}
         \caption{The Map-Hub as seen by the primary (secondary marked with a red arrow}
         \label{fig:map:primary}
    \end{subfigure}
    \begin{subfigure}[b]{0.47\textwidth}
        \centering
         \includegraphics[width=\textwidth]{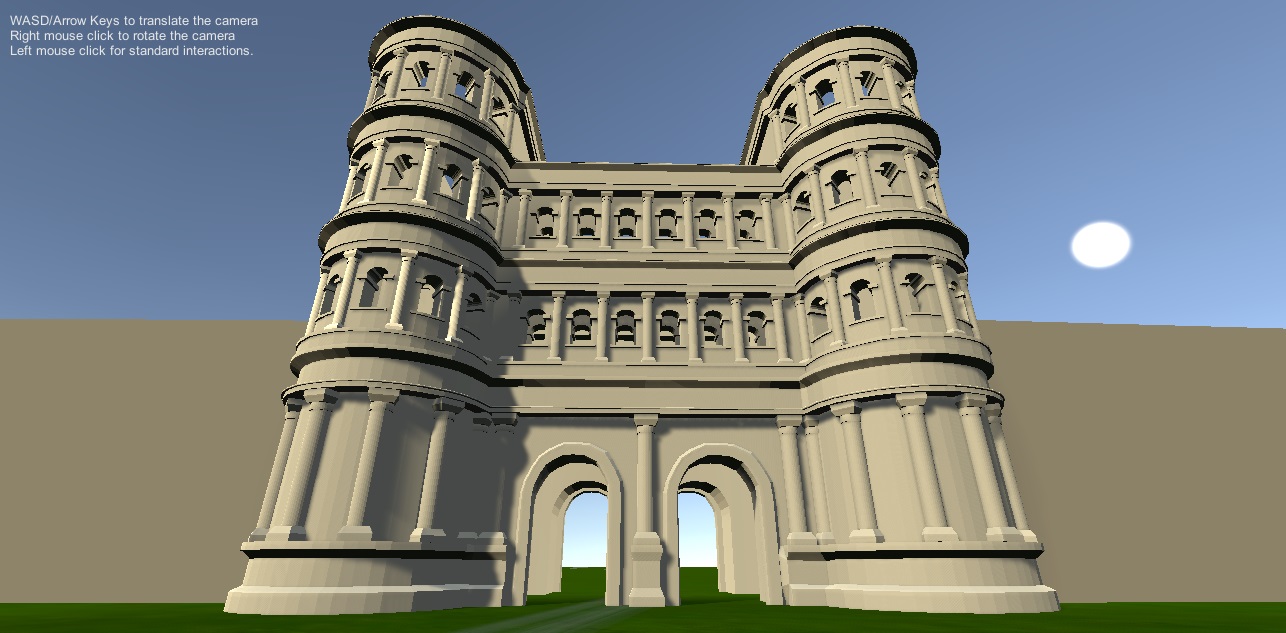}
         \caption{The corresponding view of the secondary}
         \label{fig:map:secondary}
    \end{subfigure}
    \caption{The Map-Hub and the corresponding view of the secondary}
    \label{fig:map}
\end{figure}
Figure \ref{fig:map:primary} shows the Map-Hub with the position of a secondary and the corresponding view of the secondary standing in front of the ''Porta Nigra'' is shown by figure \ref{fig:map:secondary}. 

Any secondary can now be picked up by using a shovel-metaphor with the ''Pickup-Shovel''. Therefore the primary can hold the Pickup-Shovel next to a secondary on the Map-Hub to pick him/her up. If this action was performed successfully, the secondary is no longer displayed on the Map-Hub but on the Pickup-Shovel. This allows the primary to move the secondary anywhere in his/her real physical environment. For example by placing the Pickup-Shovel on his desk the primary can now view and interact with the secondary while sitting in front of his desk, instead of standing in front of the Map-Hub. Furthermore, the virtual environment of the secondary is now also displayed to the primary (DR 1.2).

\begin{figure}
    \centering
    \includegraphics[width=.47\textwidth]{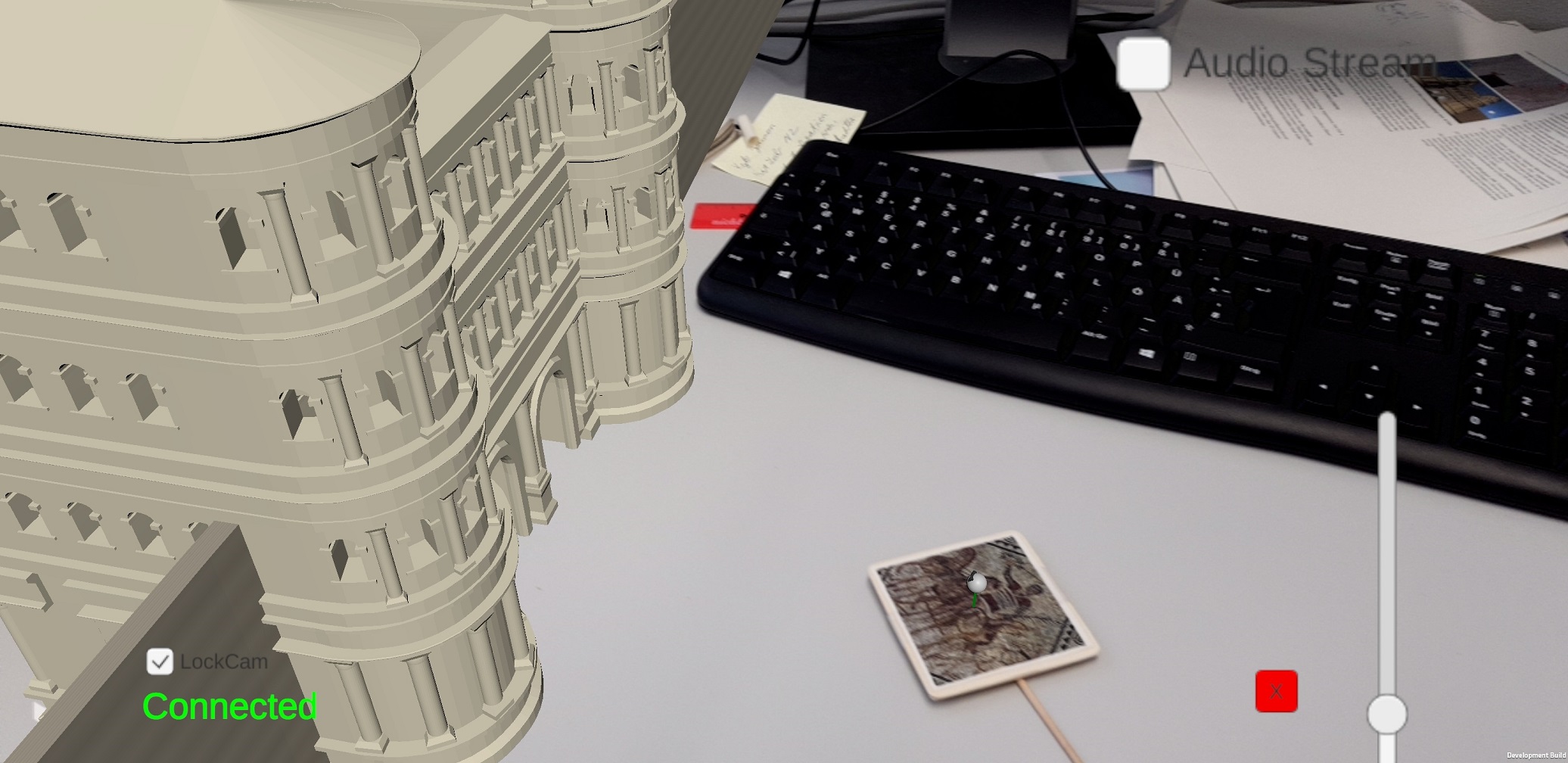}
    \caption{Interface for primary while interacting with a secondary}
    \label{fig:pickup}
\end{figure}

A voice chat can now be started by the primary via a toggle-button on his/her UI, so that both users can communicate with each other (DR 1.1). 
To keep an overview of the environment the primary can adjust the scale via a slider. If the primary wants to have an overview of the surroundings he/she can scale the environment down and if the wants to interact with a smaller object he/she can scale the environment up to allow more precise interactions. By default the primaries position remains on the center of the pickup-shovel. Therefore if the secondary moves around in his/her environment on the primaries view the environment is moving rather than the secondary. As this allows the primary to keep focus on the secondary, this mode makes it difficult to interact with the environment while the secondary is moving. Therefore, the primary can change the camera-mode to remain relative to the environment rather than following the secondaries position with an toggle-button in the interface. This allows the primary to interact with the secondary regardless of his/her position in the virtual environment (DR 3.2).
Figure \ref{fig:pickup} shows the primaries interface when interacting with the secondary. The toggle-button ''Audio Stream'' on the top right enables or disables the voice chat, the toggle-button ''LockCam'' on the bottom left changes the camera mode and the slider on the bottom left allows changing the scale. The red ''X''-button removes the secondary from the Pickup-Shovel and places him/her back into the Map-Hub, so that the primary can pickup another secondary. The secondary currently does not have any indication if he/she is currently picked up or at the Map-Hub. This could change in future work, as an indication could possibly change the secondaries behaviour and motivation to explore. 

To allow interaction with the surroundings the environment contains ''objects of interest'' (OOI). By tapping on one of these OOI, on the smartphone a context menu appears that enables various interactions. These interactions vary depending on the OOI selected, as they can be turned on and off and edited for every OOI beforehand. Current implemented interactions are:
\begin{itemize}
    \item Display Text
    \item Display Video
    \item Scale
    \item Highlight
    \item Change
    \item Lock
\end{itemize}
The first two interactions are for displaying additional information about the OOI (DR 2.2). When clicking on either the ''Text'' or the ''Video''button, a text-box or a video screen appears in front of the secondary. The text and the video must also be set beforehand.
With the ''Scale'' interaction enabled, the primary can adjust the scale of the OOI via a slider.
''Highlight'' draws an outline around the OOI for both users and displays an arrow to the secondary pointing towards the OOI. This arrow gets removed when the secondary is next to the OOI and is looking towards it. This interaction can be used by the primary to steer the focus of the secondary (DR 1.3).
The ''Change'' interaction allows the primary to change the OOI into another OOI via a dropdown menu

\begin{figure}
    \centering
    \begin{subfigure}[b]{0.47\textwidth}
        \centering
         \includegraphics[width=\textwidth]{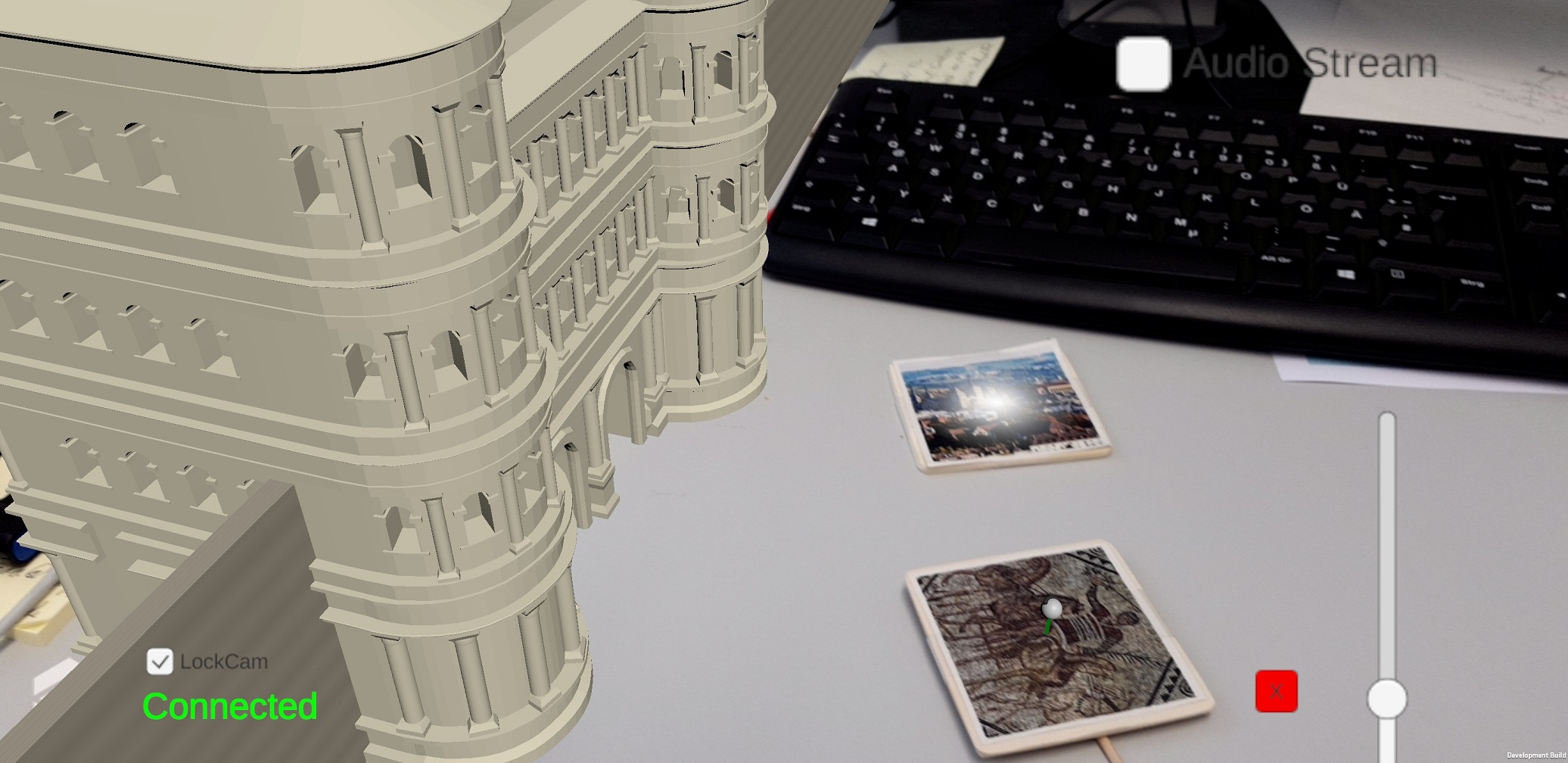}
         \caption{Tangible without an attached OOI}
         \label{fig:tangible:empty}
    \end{subfigure}
    \begin{subfigure}[b]{0.47\textwidth}
        \centering
         \includegraphics[width=\textwidth]{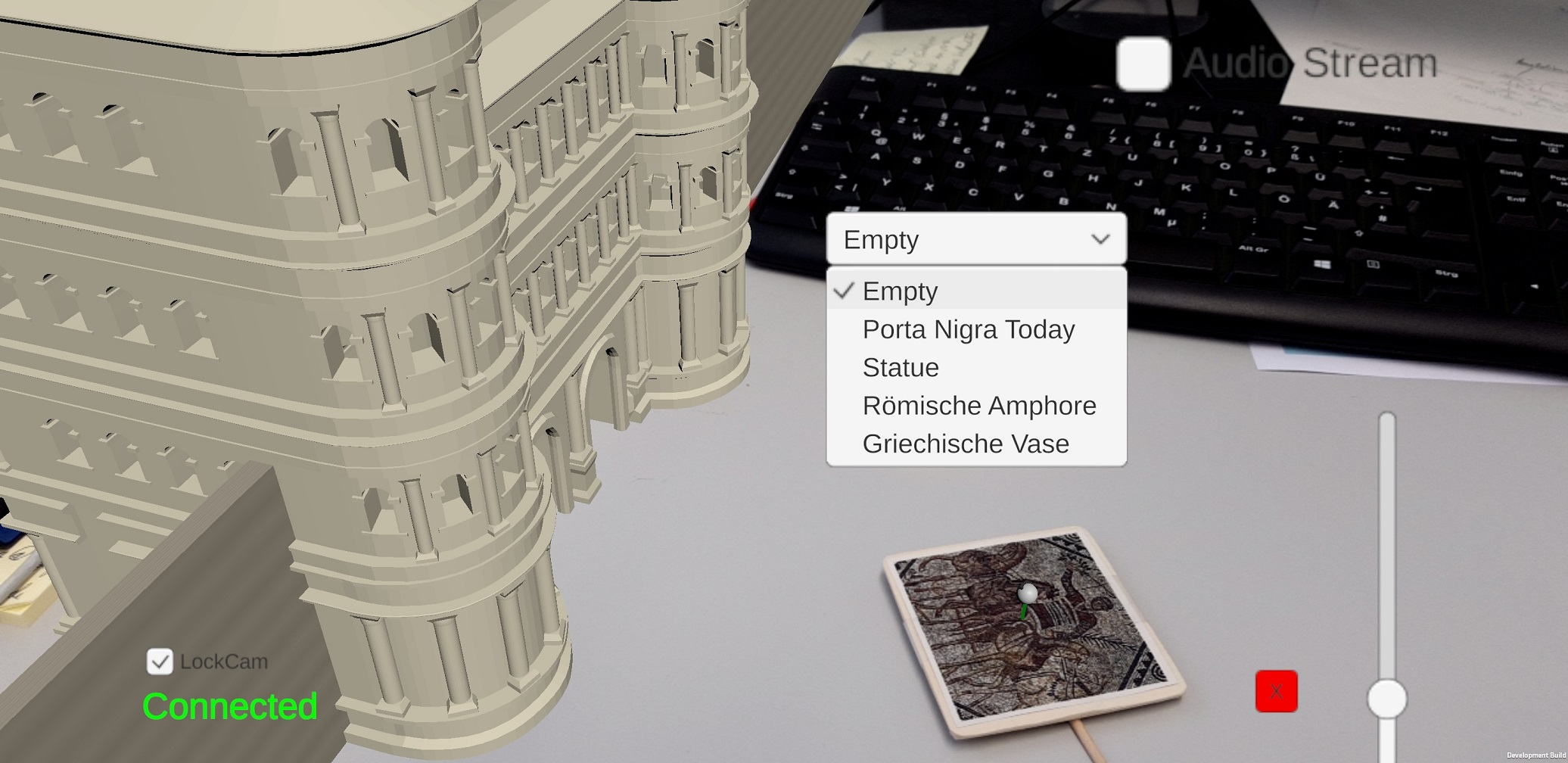}
         \caption{The dropdown menu to select a new OOI}
         \label{fig:tangible:dropdown}
    \end{subfigure}
    \begin{subfigure}[b]{0.47\textwidth}
        \centering
         \includegraphics[width=\textwidth]{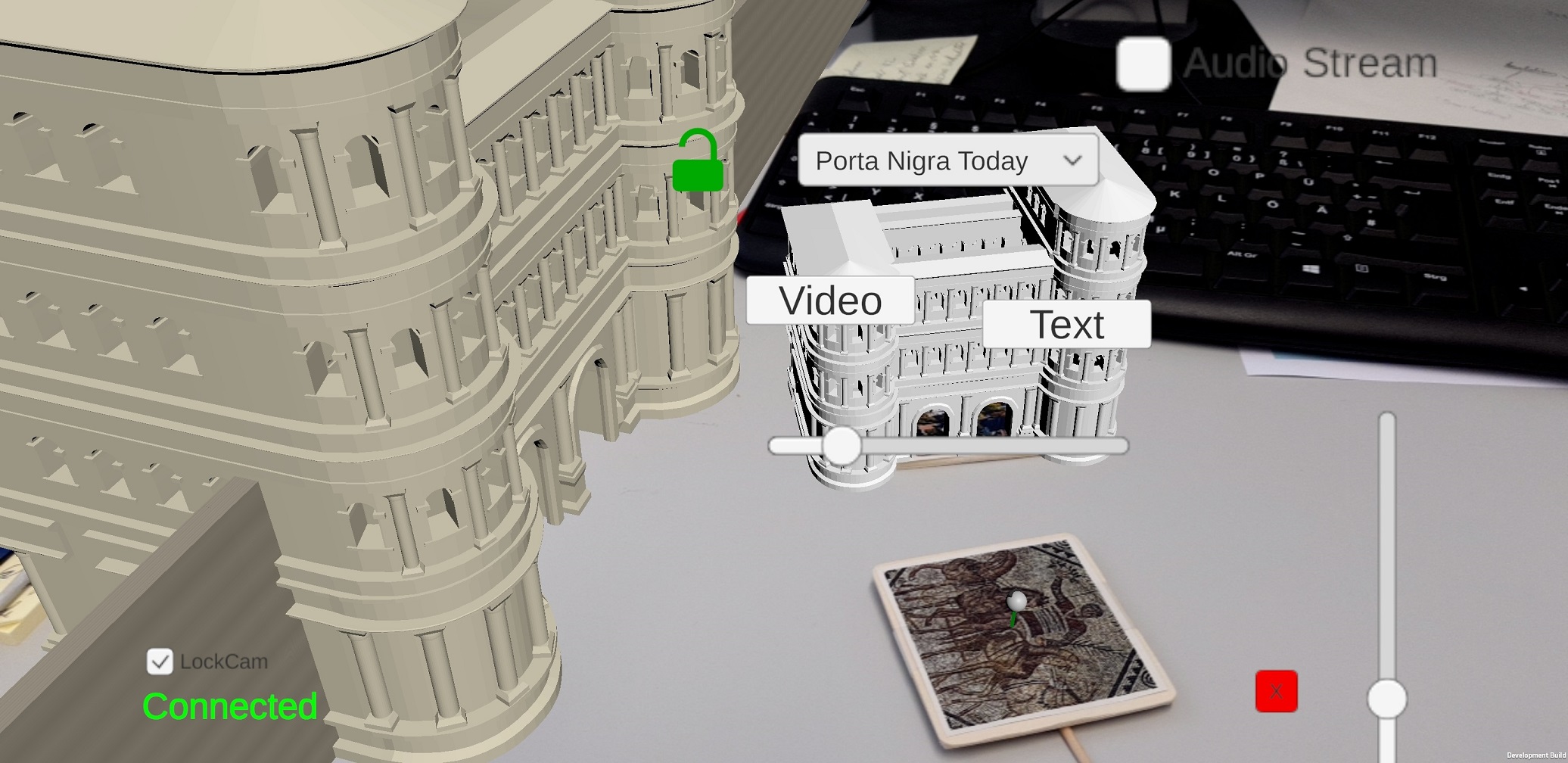}
         \caption{The context menu of the new OOI}
         \label{fig:tangible:menu}
    \end{subfigure}
    \caption{Workflow of the Tagibles}
    \label{fig:tangible}
\end{figure}

To let the primary create new OOIs for example to create context related OOIs to substantiate facts, Augmentix uses Tangibles. As seen in figure \ref{fig:tangible:empty} when placed within the view port of the smartphones camera, the tangible gets highlighted on the screen and the primary can tap on the tangible to display a dropdown menu. Figure \ref{fig:tangible:dropdown} shows the dropdown menu with the list of all possible OOIs the primary can create. By selecting one of the corresponding OOI, it gets created in the CVE, thus visible for both users. If the primary now physical moves the tangible around the primary, the created OOI now moves in the secondaries environment simultaneously. Therefore the primary can now place new OOIs next to other OOIs to compare these two.

Figure \ref{fig:tangible} shows this workflow of creating a new OOI via a tangible. 

With the last interaction ''Lock'', the primary can lock the position of the OOI, so that it does not longer follow the movement of the tangible. This is done via a click on the lock-symbol. The primary then can use the tangible to add another OOI. A locked OOI can also then be removed by a red trashcan-button. If an empty tangible is held next to an OOI it can be attached back onto the tangible by clicking the lock symbol again. Figure \ref{fig:tangible:menu} does show the OOI "Porta Nigra" currently attached to an tangible with the interactions ''Display Text'', ''Display Video'', ''Scale'', ''Change'' and ''Lock'' enabled. With the tangibles and the implemented interactions the primary can interact and manipulate the CVE and therefore DR 2.1 and DR 2.3 are met. A summary and overview of the solutions for the DRs are given in table \ref{tab:solutions}.

\begin{table}[]
    \centering
    \begin{tabular}{c|l}
        Design Requirements & Solutions \\ 
        \hhline{=|=}
        DR 1.1 & Voice chat \\
        DR 1.2 & Pickup-Shovel \\ 
        DR 1.3 & Highlight-Interaction \\ 
        \hline
        DR 2.1 & Interactions and OOIs \\ 
        DR 2.2 & Display Text/Video \\ 
        DR 2.3 & Map-Hub, Pickup-Shovel \\ 
        ~       & and Tangibles \\
        DR 2.4 & Map-Hub \\ 
        \hline
        DR 3.1 & Teleport and POI-Teleport \\ 
        DR 3.2 & Multiple Camera modes
    \end{tabular}
    \caption{Solutions to the design requirements}
    \label{tab:solutions}
\end{table}

With this proposed workflow all DRs are met. The next section does cover the technical implementation of this workflow.

\section{Implementation}
\label{sec:implementation}

This system was implemented using the ''Unity'' 3d game engine \cite{unitygameengine}. Unity allows to create 3d games for various platforms like mobile devices (android, iOS) or desktop pcs (windows, linux, macOS). 

\subsection{Augmented reality}
\begin{figure}
    \centering
    \includegraphics[width=0.3\textwidth]{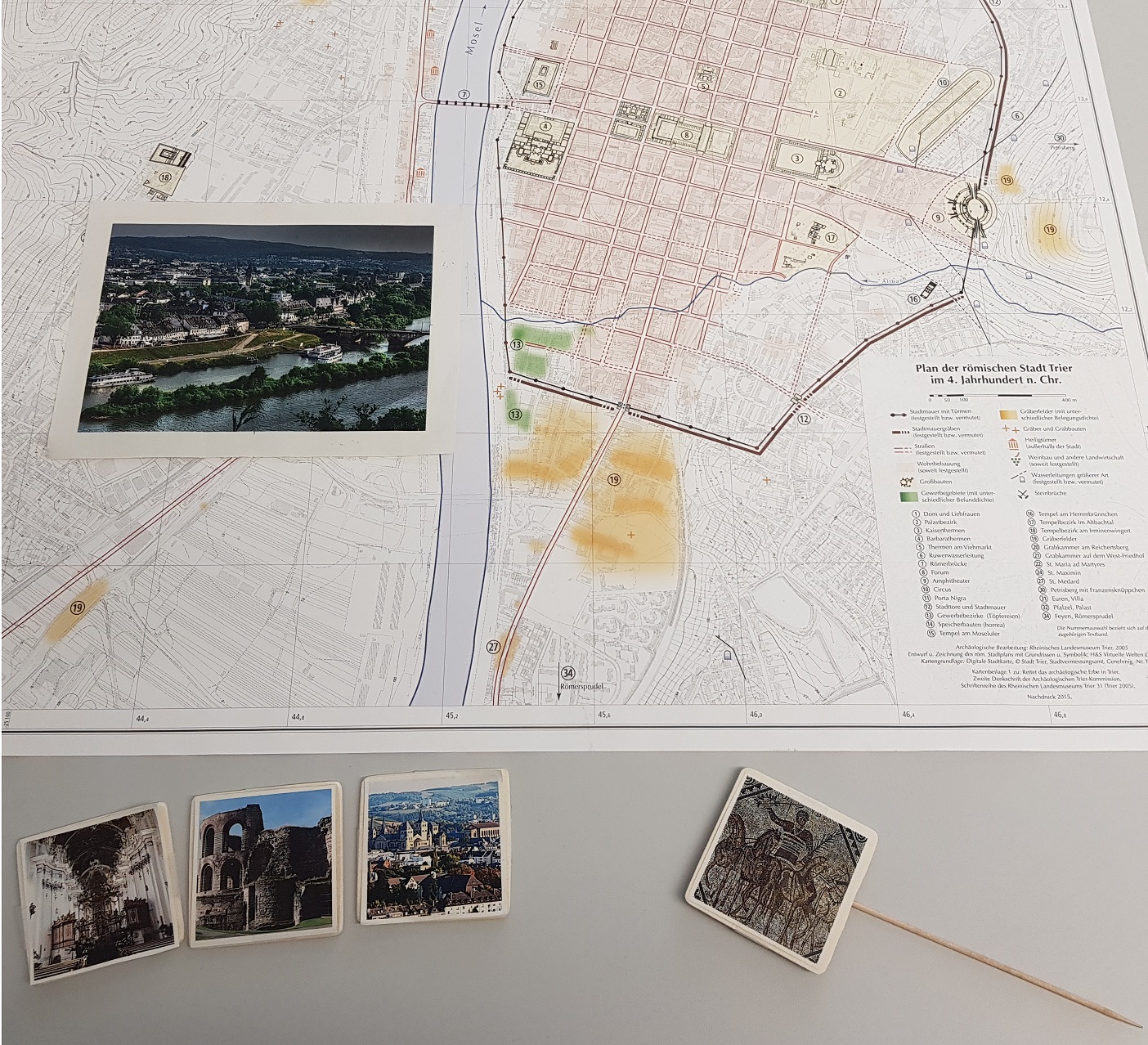}
    \caption{The physical props with their markers attached}
    \label{fig:entities}
\end{figure}

The augmented reality part of Augmentix is implemented with ''Vuforia'' \cite{vuforia} and therefore these definitions only focus on marker-based tracking. An ''augmentation entity'' describes an entity which is in a CVE used for AR. This entity consists of two major parts: a physical prop and a virtual representation in unity. The physical prop gets tracked in the real physical environment and its position and rotation gets mapped onto the virtual representation to let the virtual representation move like the physical prop does. In Augmentix the physical props are just images printed out, also called ''Markers''. The virtual representation of these markers are called ''Image Targets'' in Vuforia. 
In Augmentix there are five augmentation entities:
\begin{itemize}
    \item The Map-Hub
    \item The Pickup-Shovel
    \item 3x Tangibles
\end{itemize}
Therefore 5 unique markers were needed. These where randomly chosen images from the internet, themed in the antic age. As Vuforia allows to test images for their ability to be used as markers. Only images with the highest rating were used to minimize tracking issues. These images were then printed and stuck onto a map of Trier (Map-Hub), as small wooden shovel (Pickup-Shovel) or three small wooden chips (Tangibles) to enable physical interactions as seen in figure \ref{fig:entities}. 

To be able to scale all the game objects of one image target a dummy game object named ''Scaler'' is created as a child of each image target. Each game object for augmentation now gets attached as a child of the Scaler. If the Scaler is now scaled or moved, all the underlying game objects are scaled or moved aswell. This can be used to create an offset of the augmentation or to scale it all at once.

To implement the Map Hub one of these images was put on top of a real map. First test to use the map itself as a marker failed, as it has to few feature points, which are needed to be a good augmentable image \cite{vuforiaFeature}. Because the position of the secondaries in the Map Hub should not be based on the position of the image but rather on the position of the map, the offset between the center of the image and the origin in the secondaries virtual environment must be considered when placing Game Objects. 

\begin{figure}
    \centering
    \includegraphics[width=0.47\textwidth]{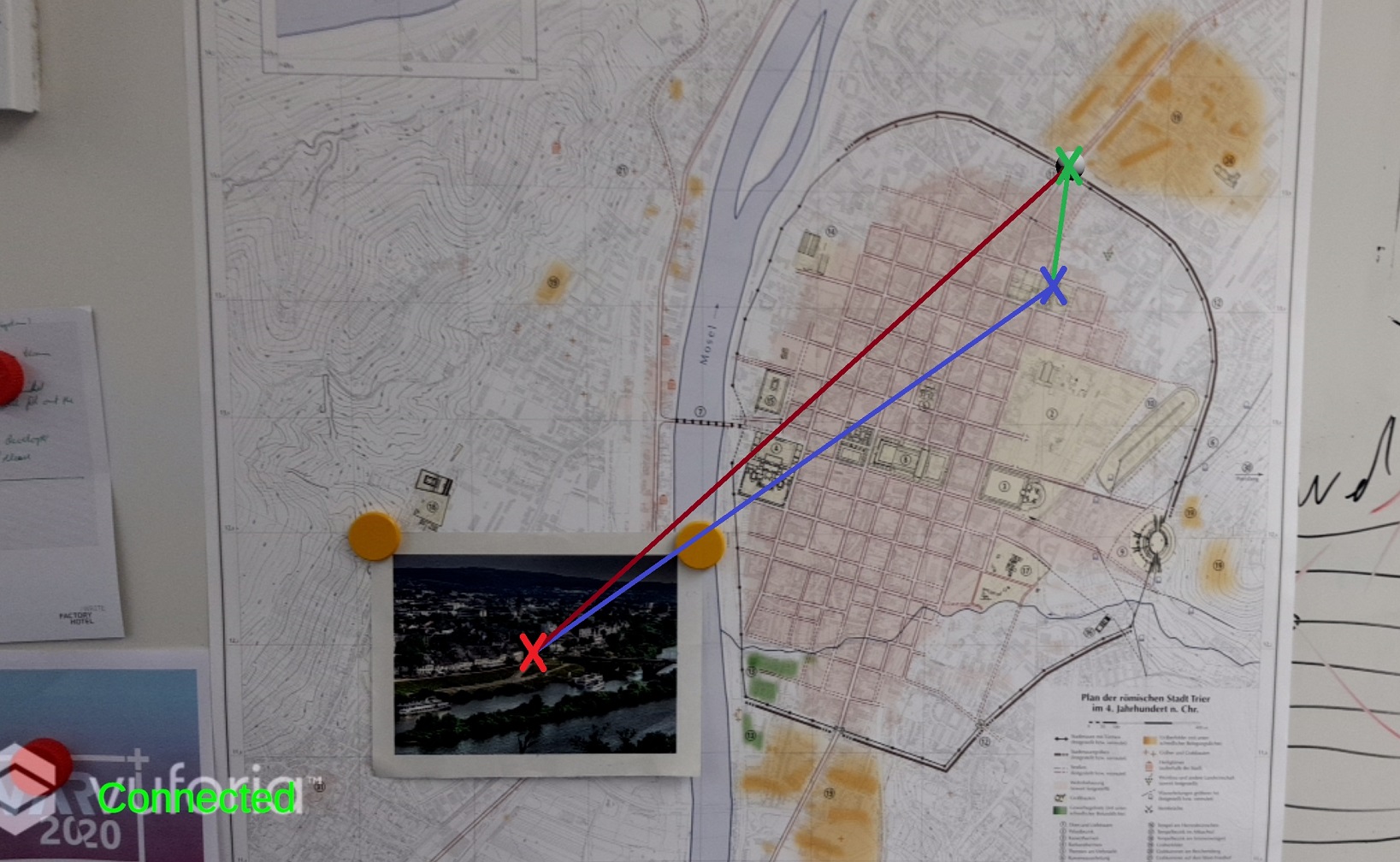}
    \caption{Visualisation of the origin of the Map Hub}
    \label{fig:map:origin}
\end{figure}

The offset can be created my moving the Scaler to the position the origin of the virtual environment of the secondary is on the map.
A visualization of this can be seen in figure \ref{fig:map:origin} where the center of the image is marked red, the origin of the virtual environment of the secondary is marked blue and the position of the player green. The brown line therefore shows the final offset from the image to the players position. The origin of the virtual environment of the secondary has been arbitrarily been chosen when the virtual representation of Trier was created.

Another image is also attached to the physical Pickup Shovel. When the image gets tracked by the primary holding the Pickup Shovel in the cameras view, the distant of the corresponding Image Target and the secondary on the Map Hub get calculated and if it is under a predefined threshold the secondary gets put as a child of the Scaler of the Pickup Shovel. Therefore the secondary is now following the Pickup Shovels position and the primary can now place the secondary somewhere else.

The Tangibles are similar to the Pickup Shovel regarding their implementation with AR. An image is attached to a wooden chip to be able to track it. When creating a OOI now as a child of the corresponding image target, the OOI is following the Tangibles position. 

\subsection{Interface}
The interface was implemented with the standard UI provided by Unity. These include common input types, like text-fields, buttons, drop-down menus and sliders. 
The interface is divided into two parts. While the ''AR-UI'' is used for any interactions with the virtual environment and can easily be implemented with the standard UI, the ''OOI-UI'' is used to interact with a selected OOI. Therefore the OOI-UI always in context to an specific OOI. To visualize this context the OOI-UIs position is always following the OOIs position. Because the OOIs position is in 3D space and the OOI-UIs position is on the 2D screen-plane a projection based on the cameras projection matrix has to be made. Figure \ref{fig:tangible:menu} does show the positioning of the OOI-UI in context to the selected OOI. Because the amount of button representing different interactions (see section \ref{sec:workflow} is dependent on the OOI selected the positions of these buttons must adjusted according this amount. Therefore the buttons get arrange in a circle around the OOI. The distance of the arc between the different buttons is dependent on their amount. The final equation for the position is:
\begin{verbatim}
x = Radius * Screen.width / 100 
    * Math.Cos(2 * i * Math.PI / Buttons.Count)
y = Radius * Screen.height / 100 
    * Math.Sin(2 * i * Math.PI / Buttons.Count)
\end{verbatim}
Where $Radius$ is a predefined constant which defines the percentual radius of the circle in proportion to the given screen space and $i$ is the index of the current Button.

\subsection{Network}
For the communication between the users the framework ''Photon'' by Exit Games was used \cite{photon}. 
This framework includes various networking-libraries for different aspect of network communication. 
We used ''PUN'' for the synchronisation of and interaction with the virtual environment and ''Voice'' for the implementation of the voice chat. 
Each game object that should be synchronized with PUN must have a ''PhotonView''-Component. 
This component handles the global indexing of these objects in the network and defines an owner of this game object. 
If various properties or components of this game object should be synchronized additional views can be added. 
These views then define which properties of a game object are synchronized and how they are synchronized. 
One game object can have multiple views handling various properties. 
Therefore a game object which synchronizes its transform and for example a health-component can do this via a transform-View and a health-View where the synchronization is handled. 
Some of these additional views, like a transform view are already provided. 
Because the original transform-view only handles the global position of objects and in the primaries virtual environment all synchronized objects are children of image targets, this view had to be adjusted to synchronize local positions rather than global positions. 
A custom transform-view was also needed to synchronize the positions of OOIs attached to a  Tangible. 
Because the OOI is a child of the tangible and therefore in another coordinate system, its local position must first be calculated into the local coordinate system of the virtual environment of the secondary. 
Thus the position of the OOI can correctly be set in the virtual environment of the secondary. This projections into different local coordinate systems are provided by Unity. 

Because all game object used must be available to all users, every game object synchronized over the Network, which is not a part of the scene from the beginning, must be defined as a prefab-asset in the ''Resource'' folder of the project. Is is especially necessary because Unity does optimize the system when build, by removing all unused assets. Therefore if an asset is not used in a scene it will get removed by unity and can't be used at runtime. An exception to this behaviour is the ''Resource'' folder. This one does get ignored by unity and every asset inside this folder is available at runtime. These assets include:
\begin{itemize}
    \item A player avatar
    \item the virtual representation of the city
    \item every OOI creatable with a Tangible
\end{itemize}
A player avatar get created for each user every time a new secondary is starting their application and synchronize their transform with each user. Additionally a instance of the virtual representation for the city gets created for the secondary to walk in and for the primary. On the primary site this representation gets hidden first and then is later used, by making it visible again, in the Pickup-Shovel to show the surroundings of the secondary to the primary. 

\section{Expert interview}
\label{sec:interview}
To validate the system and the workflow we conducted an expert interview. In this interview six computer scientist separately were first introduced into the concept and the goal of this system. Then they could freely try the secondaries site of this system to get familiar with the secondaries virtual environment and interactions. After that they got introduced into the workflow and the system of the primary and then could freely try the primaries site and get used to the workflow and all interactions provided by Augmentix. Subsequently they were asked to perform specific tasks to guide a student through a staged virtual city tour. These tasks can be found in the attachments. The secondary was here played by one of the authors of this paper and intentionally forced the expert to use all provided interactions. After the tasks were successfully completed the expert was asked a eight questions about his/her experience. These questions aimed at the functionality of the system and the advantages and disadvantages the usage of AR. These questions can be found in the attachments. The answers to these questions were anonymously recorded, transcribed and translated from german to english afterwards. 

\begin{table*}[ht]
    \centering
    \begin{tabular}{cp{0.7\linewidth}l}
     Usage of AR & E2: "... can better assess the behaviour of the user, because I am more immersed" & I.a \\ 
     & E3: "[The interaction] is very natural ... therefore i think for normal city tour guides [this system] is better, than a desktop environment ..." & I.b \\ 
     & E4: "to place the [objects] in the view of the player was quite fiddly ... i think this will be easier with a mouse" & I.c \\ 
     \hline
     Workflow & E1: "... it was cool ... nothing to complain." & II.a \\ 
     & E5: "I think it is difficult with two planes (work spaces) ... you could project the Map-Hub onto the desk" & II.b \\ 
     & E2: "I like the metaphor, but i don't think it does scale well [with multiple students]." & II.c\\
     \hline
     Interaction & E1: ".. can not put tangibles very close to each other in the real world." & III.a \\ 
     & E2: "It was confusing ... to change the interaction space." & III.b \\ 
     & E5: "I always tried to click on the [physical] tangibles. & III.c \\ 
     & E1: "[The interface] was quite intuitive ..." & III.d \\ 
     \hline
     Features & E3: "I think ... a rotation ... could be very helpful and make [the guidance] easier. " & IV.a \\ 
     & E2: "Pointing would be a nice feature, like 'Hey, look here' & IV.b \\ 
     & E5: "What could be more intuitive is if the scaling could be done via a [multitouch] gesture & IV.c \\ 
     \hline
     Technical Aspects & E2: "I see a big disadvantage in the mobile phone ..." & V.a \\ 
     & E3: "Depending on the amount of students ... i think two [free] hands are clearly superior." & V.b \\ 
     & E1: "The buttons are really close to each other" & V.c \\ 
     & E6: "I couldn't really see where you were facing" & V.d \\ 
\end{tabular}
    \caption{Expert feedback}
    \label{tab:expert}
\end{table*}

While all experts think that the usage of AR as a primary is very intuitive, immersive and motivational $^{I.a,I.b}$, they also think that at the current stage this system is to inconvenient to use for an efficient and long term usage$^{I.c}$. One major factor for this inconvenience was the usage of a  smartphone $^{V.b,V.c}$. To view the virtual content the smartphone must be always held in view and therefore all the interactions can only be performed with a single hand. As this flaw was suspected and therefore two questions specifically focused on this topic, the expert interview confirmed that the smartphone had an significant negative impact on the performance and experience of the primary.   
However because of the intuitive interactions three experts stated that they see a big advantages for users as primary who are no experts in computer systems in comparison to a potential system using a common desktop environment $^{I.b}$. But on the other hand a system using a desktop environment would probably be more efficient for users once they get familiar with this system. To counter these problems five experts suggested the use of an AR HMD as this would allow the primary to use both hands for interactions. Three experts further suggested a stand, like a tripod, for the smartphone if no HMD is available. 

The workflow of the Map-Hub and the Pickup-Shovel was according to all experts easy to understand and convenient to use $^{II.a,II.c}$. However all experts think that this workflow does not scale well with multiple secondaries$^{II.c}$. While the pickup-metaphor was easy to understand it would be too tricky and imprecise to use and the overall overview would suffer when handling multiple secondaries. Two experts also stated that they think merging the Map-Hub and the Pickup-Shovel into one plane could improve this workflow, as both work spaces are then always in the field of view of the primary$^{II.b}$.

The interface and the interactions were received well by all experts. The interface was clear and easy to understand. Only the buttons sometimes overlapped and then were hard to click on$^{V.c}$. No expert missed an interaction method while performing the tasks. 
The usage of tangibles were also described as intuitive and easy, merely the physical prop was perceived as too clunky to perform precise interactions with them $^{III.a}$. Furthermore two experts were bothered by the repeating context switch between the physical tangibles and the interface of the smartphone$^{III.b, III.c}$. Further two experts stated that the secondaries view direction was not clear enough$^{V.d}$.
As potential new features three experts proposed a 3D pointing interaction for the primary to highlight not only objects but to point on specific areas in the environment, like a laser pointer$^{IV.b}$. Furthermore, three experts proposed a forced rotation on the secondary to guide his/her focus additionally to the already implemented highlight interaction$^{IV.a}$. A support of multi-touch gestures, like a pinch gesture, were proposed by five experts, as it would use the already known gestures used my common smartphone applications$^{IV.c}$. Furthermore three experts had technical problems, as the system did not always detect the images for the Image Targets, because they unintentional blocked the view of these images while interacting with the tangibles. 

Because in the interview the primary and the secondary were in the same room the usage of the provided voice chat was not needed. Nevertheless all experts stated that a voice chat would probably be sufficient to communicate with the secondary in a distributed setup and don't see any necessity for additional communication channels. Two experts stated however that the secondary could possible benefit from a video chat, because he could be feel more engaged with the primary when seeing him via a video chat.

\section{Discussion \& Conclusion}
\label{sec:conclusion}
While the expert interview showed, that the usage of AR as a primary does potentially provide a intuitive and immersive experience, in the current stage it does not enhance the workflow of the primary in comparison to a potential system using a common desktop environment. One major drawback was the usage of a common smartphone as an input and display device. Therefore the technical implementation of AR has a huge impact on the experience of the primary. Furthermore the proposed workflow was well received in terms of design and usability, but, according to the expert interview, will not scale well with multiple students. 
This leads to the conclusion, that the workflow must not only be designed in terms of usability and complexity, but must be efficient and scaleable, aswell. Therefore the current workflow does not provide the desired enhanced experience for the primary. 
The interface was well received and the interaction methods were easy to understand, therefore only minor tweaks are necessary to optimise these aspects. But the interface and the interaction methods heavily depend on the implementation of AR. In this first prototype these were designed for a smartphone, but if AR is implemented with a HMD the interface and the interaction methods must be completely redesigned. 
In terms of communication channels the provided voice chat was stated as sufficient and therefore no additional channels are needed.  

These results lead to the conclusion that this first prototype of Augmentix is not sufficient enough to tackle the two research questions identified in section \ref{sec:introduction}. While the experts stated that this system is easy, intuitive and immersive, no positive aspects on performance were identified. But we think that all the negative aspects can be addressed and a positive aspects in performance can be achieved by the usage of AR. 

\section{Future Work}
\label{sec:future}
In potential future work first the negative technical aspects should be addressed. The authors suggest research which implements AR with a HMD to allow a more immersive and intuitive interaction. Further the tangibles could be upgraded to allow more physical interactions. A physical button or switch could minimize the need of context switches, which had a negative impact in this first prototype of Augmentix. 

In addition a more scalable and efficient workflow must be designed. Though the current workflow is easy and intuitive it was stated as not scalable and efficient by the experts. To achieve significant results the potential new workflow must be validated in a study with multiple secondaries in one sitting. 

To finally validate such a system using AR to enhance the primaries workflow, a desktop application for guiding a secondary through a virtual city must be designed. As such a system is needed to verify the results claimed about a system using AR.







\bibliographystyle{ACM-Reference-Format}
\bibliography{sample-base}
\end{document}